\documentclass{iau} 
\usepackage{graphicx}

\title[Sun's polar magnetic field] 
{The Sun's polar magnetic field: datasets, proxies and theoretical issues}

\author[A.\ R.\ Choudhuri]   
{Arnab Rai Choudhuri$^1$}

\affiliation{$^1$Department of Physics, Indian Institute of Science, \\ Bangalore -- 560012, India
 \\ email: {\tt arnab@iisc.ac.in} }

\pubyear{2017}
\volume{340}  
\jname{Long-Term Datasets for the Understanding of Solar and Stellar Magnetic Cycles}
\editors{A.C. Editor, B.D. Editor \& C.E. Editor, eds.}
\begin{document}

\maketitle

\begin{abstract}

The polar magnetic field of the Sun is a manifestation of certain aspects of the dynamo
process and is a good precursor for predicting a sunspot cycle before its onset. Although
actual synoptic measurements of this field exist only from the mid-1970s, it has now
been possible to determine its evolution from the beginning of the twentieth century with
the help of various proxies. The recently developed 3D kinematic dynamo model can study
the build-up of the Sun's polar magnetic field more realistically than the earlier surface
flux transport model.


\end{abstract}

\firstsection 
\section{Introduction}

The Sun's polar magnetic field was discovered by Babcock \& Babcock (1955).  Within a few 
years of this discovery, Babcock (1959) found that the polar field reversed during the sunspot
maximum.  The low-resolution magnetograms of that era indicated the polar magnetic field to be 
of the order of a few gauss. One important question was whether this was a truly weak field pervading
the whole polar region or whether the field existed in the form of fibril flux tubes which were
not resolved by the low-resolution magnetograms. This question could only be settled
with observations from space due to Hinode, which showed fibril magnetic field configurations
even in the polar regions with about 1000 G magnetic fields in their interiors 
(Tsuneta et al.\ 2008).  It may be
mentioned that, a few years before this discovery, Choudhuri (2003) had predicted on the basis
of some dynamo arguments that the polar magnetic field would have such a configuration.

Systematic synoptic data of the polar magnetic field started being recorded 
from the mid-1970s---mainly from the Wilcox Solar Observatory. From such data, it is now clear
that the polar field varies with the sunspot cycle in a roughly periodic fashion---changing 
its sign around the sunspot maximum and becoming strongest during the sunspot minimum. It also
appears that there is a correlation between the maximum strength of the polar field during a sunspot
minimum and the strength of the next sunspot cycle. If there is indeed such a correlation, then
one can use the strength of the polar field during a sunspot minimum to predict the next cycle
(Svalgaard, Cliver \& Kamide 2005; Schatten 2005). One question before us is whether we expect
such a correlation on the basis of theoretical dynamo models.  The other question is whether
we can extend the dataset of the polar field for a few decades before the 1970s by using other
proxies so that we can look for this correlation in a more extended dataset.  
These issues are discussed in the next two sections. Then Section~4 will discuss some
theoretical issues connected with the build-up of the polar field.

\section{The significance of the polar field in the flux transport dynamo}

In the currently popular flux transport dynamo models of the Sun 
(Choudhuri 2011, 2014; Charbonneau 2014; Karak et al.\ 2014), it is assumed that 
the poloidal field is produced from the decay of tilted active regions, as first
envisaged by Babcock (1961) and Leighton (1969).  This poloidal field is then transported
poleward by the meridional circulation of the Sun to build up the polar field at the
end of the sunspot cycle. At the same time, the poloidal field diffuses to the tachocline
across the convection zone to act as the seed for the next sunspot cycle. Choudhuri,
Chatterjee \& Jiang (2007) had argued that the main source of irregularities in the
solar cycle is the randomness in the Babcock--Leighton process arising out of the
scatter of the tilt angles around Joy's law (Longcope \& Choudhuri 2002). Jiang,
Chatterjee \& Choudhuri (2007) have given the following explanation of how the correlation
between the polar field at the end of the sunspot cycle and the next sunspot cycle
arises. Suppose the randomness in the Babcock--Leighton process has made the poloidal
field produced in a cycle stronger than the average.  Then we expect both polar
field at the end of the cycle and the seed for the next cycle to be stronger.  On the
other hand, if the poloidal field produced in a cycle is weaker, then both of these
things would be weaker.  We thus understand how the observed correlation arises. This
discussion makes it clear that the polar field is an important manifestation of the
flux transport dynamo model.

\section{Proxies for the polar field in the past}

Since it appears that the polar field is a good precursor for predicting the sunspot
cycle and, because of its importance for the dynamo process, we would like to obtain
information about its evolution before the 1970s when direct measurements began. The number
of polar faculae in Mount Wilson white light plates has been found to be a good proxy of the solar
polar field (Sheeley 1991; Munoz-Jaramillo et al.\ 2012).  From the polar faculae counts,
it has been possible to reconstruct how the polar field might have varied from the
beginning of the twentieth century. Apart from the Mount Wilson Observatory, the other observatory where
solar plates have been taken during a large part of the twentieth century is the Kodaikanal
Observatory. From two kinds of plates taken at the Kodaikanal Observatory, two different
proxies of the polar field have been found.  From the Kodaikanal H$\alpha$ plates, one
finds positions of prominences indicating magnetic neutral lines and, from their positions,
the history of polar field evolution has been reconstructed (Makarov et al.\ 2001). Recently,
Priyal et al.\ (2014) discovered that the network bright points in Kodaikanal Ca K plates provide
another good proxy for the polar field.
 
With various proxies for the polar field available at earlier times, it is worthwhile to check
whether the correlation between the polar field during a sunspot
minimum and the strength of the next sunspot cycle can be seen in the earlier data before
the 1970s.  Such correlation seems to exist even in the earlier data (Jiang, Chatterjee \&
Choudhuri 2007; Goel \& Choudhuri 2009). By incorporating the polar field data in theoretical
dynamo models, Goel \& Choudhuri (2009) could model various aspects of sunspot cycles, including
the hemispheric asymmetry.

\section{The build-up of the polar field from the decay of active regions}
 
At last, we come to the issue of constructing realistic models of how the polar field builds
up from the poloidal field arising out of the decay of bipolar sunspots by the Babcock-Leighton
process---due to the combined effects of diffusion, differential rotation and meridional
circulation. This is done in the surface flux transport model, which started being developed
from the pioneering work of Wang, Nash \& Sheeley (1989) and has been extended by the group
at MPS in the past few years.  In this model, the radial component $B_r$ of the magnetic field
is treated as a scalar and its evolution is studied on the two-dimensional solar surface spanned
by $(\theta, \phi)$ coordinates. In spite of its many successes and its great historical 
importance, the surface flux transport model has two limitations: (i) the vectorial nature of
the magnetic field is not included; and (ii) the radial component of the meridional circulation
is not included, so that the subduction of the magnetic field in the polar regions due to the
sinking meridional circulation there is not handled properly.
  
\begin{figure}
\begin{center}
\includegraphics[width=4in]{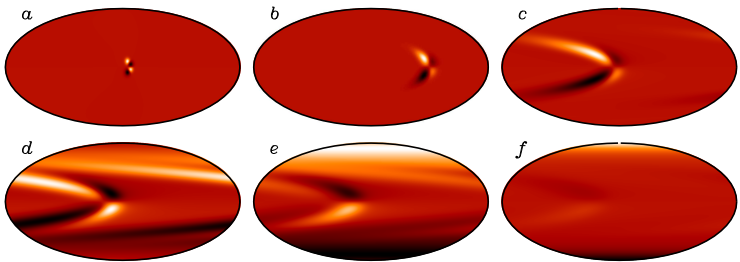} 
 \caption{Evolution of the magnetic field on the solar surface starting from two tilted
  sunspot pairs placed symmetrically in the two hemispheres. From Hazra, Choudhuri \& Miesch (2017).}
   \label{fig1}
\end{center}
\end{figure}

\begin{figure}
\begin{center}
\includegraphics[width=3.5in]{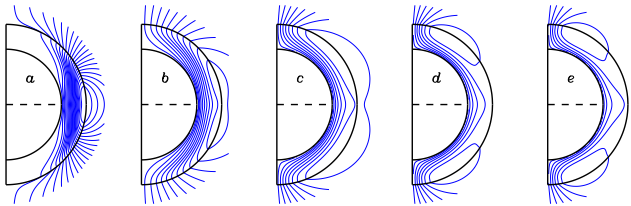} 
\caption{Evolution of the longitudinally averaged poloidal field for the case shown in
Figure~1. From Hazra, Choudhuri \& Miesch (2017).}
   \label{fig2}
\end{center}
\end{figure}

The recently developed 3D kinematic dynamo model can be used for studying the build-up of the
polar field more realistically and in a manner free from the above-mentioned two limitations
(Hazra, Choudhuri \& Miesch 2017).  This model is kinematic in the sense that the flows like the 
differential rotation and the meridional circulation are put by hand, but the magnetic field
is treated in 3D. Let us consider one example which brings out the difference of this model
from the surface flux transport model in a rather dramatic manner. Suppose we put two tilted
bipolar sunspots in the two hemispheres symmetrically.  Figure~1 shows how the magnetic field
on the solar surface will evolve according to the 3D kinematic dynamo model and eventually build
up the polar field.  Figure~2 shows the evolution of the longitudinally averaged 
poloidal field lines during this process.
In Figure~2(c), we see that the magnetic field at the surface is almost concentrated in the polar 
regions. However, because of the 3D nature of the magnetic field, further stretching of field
lines due to the effect of the meridional circulation causes $B_r$ of opposite polarity to appear
at low latitudes, as seen in Figure~2(d). We thus have magnetic bubbles which sink underneath
the surface with the sinking meridional circulation in the polar regions, thereby making the
magnetic field to disappear eventually from the surface.

Exactly the same example worked out by using the surface flux transport model yields a
totally different result (Jiang, Cameron \& Sch\"ussler 2014). Once the magnetic field is
concentrated in the polar region, an asymptotically steady state is reached,
with outward diffusion being balanced by the inward meridional circulation.  If there are no
sources at the lower latitudes, then $B_r$ of opposite polarity cannot develop at the lower
latitudes, as in the case of Figure~2(d). Since the sinking of the meridional circulation
in the polar region is not considered, the magnetic field cannot sink underneath the surface.
Jiang, Cameron \& Sch\"ussler (2014) found an asymptotically steady magnetic field. The lesson
is clear.  Although the surface flux transport model gives very good results at low latitudes,
its results at high latitudes should be treated with caution, as it does not include important
physical effects.  The build-up of the polar field can be treated more realistically in the 3D
kinematic dynamo model.

Sometimes one finds large active regions not obeying Hale's law.  Jiang, Cameron \& Sch\"ussler (2015)
argued that a few large anti-Hale regions during the cycle~24 might have been responsible for
the weak polar field at the end of the cycle.  Hazra, Choudhuri \& Miesch (2017) used the 3D
kinematic dynamo model to study the effect of large anti-Hale regions and found that they have
some effect when they appear at high latitudes in the mid-phase of the cycle, but the effect
was not very dramatic.  The 3D kinematic dynamo model is particularly suited to study how the
scatter around Joy's law can cause fluctuations in the polar field (Karak \& Miesch 2017).



{\it Acknowledgments.} My
research is supported by DST through 
a J.C.\ Bose Fellowship.

\def\apj{{\it ApJ}}
\def\mnras{{\it MNRAS}}
\def\sol{{\it Solar Phys.}}
\def\aa{{\it A\&A}}
\def\gafd{{\it Geophys.\ Astrophys.\ Fluid Dyn.}}

\end{document}